\documentclass[aps,prb,twocolumn,showpacs,floatfix,superscriptaddress]{revtex4}

\usepackage{amsmath}
\usepackage{graphicx}
\usepackage{dcolumn}
\usepackage{bm}

\begin{document}

\title{Dynamics of two interacting electrons in a one-dimensional crystal with impurities}
\author{P.\ E.\ de Brito}
\author{E.\ S.\ Rodrigues}
\affiliation{Universidade Cat\'{o}lica de Bras\'{\i}lia,
Departamento de F\'{\i}sica, Campus \'{A}guas Claras, 72030-070
Bras\'{\i}lia - DF, Brazil}
\author{H.\ N.\ Nazareno}
\affiliation{International Center of Condensed Matter Physics,
Universidade de Bras\'{\i}lia, P.O. Box 04513, 70910-900
Bras\'{\i}lia - DF, Brazil}

\begin{abstract}
We investigated the role that the electron-electron interaction
plays on the propagating properties of wave packets in a
one-dimensional crystal with impurities. We considered two
interacting particles with opposite spins in a band, where we
treated their interaction along the Hubbard model. We have
obtained the density of states of the crystal for different values
of the interaction term, as well as solved the dynamical
Schr\"{o}dinger equation by varying the initial conditions. We
have introduced a method through which we were able to follow the
time evolution of the wave packets for both spins showed in
three-dimensional plots, and have evaluated, for each particle,
the corresponding $MSD^{\prime }s $ and the centroids as function
of time . These measurements allow us to determine the influence
of the interaction on dynamical properties. We discussed the
combined effect that the extension of the initial wave packets and
the interaction strength have on propagating properties. Under
certain conditions we obtained an entanglement of the two packets
associated with both spins that takes place in a small region of
the lattice.
\end{abstract}


\maketitle

\section{Introduction}

According to the scaling theory of localization developed by Abrahams et al~%
\cite{ab} in low dimensional systems (one and two dimensions), any degree of
disorder will prevent the appearance of a metallic phase. Moreover, former
experiments done with two-dimensional electron systems fabricated on
semiconductor surfaces showed a logarithmic increase of resistivity while
lowering the temperature~\cite{do,bi1,bi2,ur}. This behavior lends support
to the scaling theory of localization, since this happens in case of a weak
electron-electron interaction. These results were in agreement with
theoretical predictions~\cite{al} that \textit{weak} electron-electron
interaction increases localization. The above experiments were done with
samples of high density of electrons, i.e. systems for which the associated
Wigner radius $r_{s}\ll 1,$ which is the ratio between the Coulomb energy to
Fermi energy.

However, such a scaling theory does not take into account the
electron-electron interaction which was lately believed to be responsible
for a metal-insulator transition, observed in several experiments performed
at zero magnetic field described below. In the metallic phase, one observes
a strong temperature dependence (a steep $d\varrho /dT>0)$ caused by the
delocalizing effects produced by the interaction between the particles.

As the density is reduced such that $r_{s}\gg 1,$ the interaction becomes
dominant, for that regime Finkelstein~\cite{fi} and Castellani et al.~\cite
{ca} predicted that for sufficient strong interactions, a 2D system should
present a conducting phase as the temperature is lowered. Since recently was
possible the fabrication of 2D samples of high quality with very small
amount of randomness, measurements were done at very low particles
densities. In this way the strongly interacting regime ( $r_{s}\gg 1$) has
become experimentally accessible. For instance for $r_{s}>10$ experiments
done on low-disordered $2D$ silicon samples demonstrated that with
increasing electron density one can cross from the insulating regime where
the resistance \textit{diverges} with decreasing temperature, to a regime
where the resistance \textit{decreases} strongly with decreasing
temperature, clearly showing metallic behavior~\cite{kr1,kr2,kr3}.

In addition to that, in an extensive numerical analysis of the
two-dimensional Anderson model with dimerised disorder, we have reported the
existence of several dynamical regimes~\cite{na02,na03}.

As far as we are concerned there is no a theoretical explanation that
describes adequately the metal-insulator transition in two-dimensional
systems, as well as the dramatic increase of the spin susceptibility in its
vicinity. Comprehensive studies of the state of the art of this intriguing
problem are presented by Abrahams et al.~\cite{ab1} and Kravchenko and
Sarachik~\cite{sar}.

In one-dimensional systems with random disorder of any intensity,
all states are exponentially localized as shown in the pioneering
work by Anderson~\cite {an} in dealing with diagonal disorder,
i.e., a model where the on-site energies are randomly distributed.
\textit{\ }On the other hand, when some correlation is included in
the model, and without considering interaction between particles,
this picture is substantially modified, given place to the
appearance of extended states and, consequently, carriers are able
to propagate through the system. Several structures with
correlated disorder show vanishing of localization, when one
considers nearest neighbors hopping. Among them one can quote the
random-dimer model that can explain the high conductivity of
polymers~\cite{du,ph,si,ti}. Another example that show correlated
disorder, responsible for particle diffusion in $1D$ is provided
by the structures where the on-site energies follow the Fibonacci
and Thue-Morse sequences~\cite{ko,os,ri,ma,br}. One should mention
also the Harper model of a quasicrystal which presents a mobility
edge when the strength of the potential equals the half-bandwidth.
Starting with a well localized particle in the lattice, as long as
the Harper potential is less than half-bandwidth, we encounter
ballistic propagation~ \cite{au,so,ya,na}. The purpose of this
work is to analyze the role the electron-electron interaction
plays on propagation in some 1D nonperiodic structures. First we
present in Sec. II the model assumed for the interaction between
two particles in a band, namely the Hubbard Hamiltonian. The two
electrons are assumed in the singlet state in order to detect the
effect of the Hubbard term, since it acts on opposite spins. We
show the density of states for different strengths of the
interaction. The dynamics tools introduced in order to
characterize the dynamical behavior are presented in Sec. III,
namely, the time evolution of the mean square displacement (MSD)
and the centroids associated with each of the particles, as well
as the construction of 3D graphs of the wave packets evolution. In
sec. IV we discuss the interplay between the strength of the
interaction U with the initial wave packet extension. In Sec. V we
present the results concerning to a 1D crystal with impurities. In
Sec. VI we present the conclusion to which we arrived in this
work.

\section{The Hubbard Hamiltonian for two electrons interacting in a band}

With the aim to study the influence of the Coulomb interaction between
carriers in a $1D$ lattice, we treat the electron-electron interaction along
the Hubbard~model \cite{jh} in a crystalline system with impurities.
As the experiments show, the
electron-electron interaction should modify the behavior of the carriers as
obtained in a one particle non-interacting scheme. In our work we introduce
two interacting particles in an otherwise empty band. The study of such a
problem have deserved a number of interesting works due to the relevant
question of what is the role of the interaction between electrons on
propagating properties in low dimensional disordered systems~\cite
{she,im,wm,js,fm,ex,ek}. We assume that by solving the present problem one
can get a better understanding of the role the interaction plays in real
systems.

We consider a one-dimensional lattice of $N$ sites with lattice
parameter $d$, for which the Hubbard Hamiltonian is:

\begin{widetext}
\begin{equation}
H\,=\,\sum_{r,s}\,\,c_{r,s}^{+}c_{r,s}\gamma
_{r}\,+\,V\,\sum_{r,s}\,\,(c_{r+1,s}^{+}c_{r,s}+c.c.)\,+\,U\,\sum_{r}\,\,
{\hat{n}}_{{r}\uparrow} {\hat{n}}_{r\downarrow }  \label{ham}
\end{equation}
\end{widetext}

\noindent were $\gamma _{r}=\varepsilon _{r}+Fedr$, $\varepsilon
_{r}$ being the on-site energy, $F$ is the intensity of a dc
electric field, $c_{r,s}^{+} $($c_{r,s}$) is the Fermi creation
(destruction) operator for an electron of spin $s$ at site $r$, $V$
is the hopping term and
${\hat{n}}_{r\uparrow}$(${\hat{n}}_{{r}\downarrow }$) is the number
operator for spin up (spin down) at site $r$. As it was said above,
in order to analyze the role the $U$ term plays on the dynamics, we
treat the case of electrons with opposite spins, the singlet.

\subsection{Energy spectrum for a crystal}

To obtain the energy spectrum for the singlet in an impurity-free
1D lattice, we solve the stationary Schr\"{o}dinger equation in
the Wannier representation where we expand the eigenfunction in
terms of the kets $|ns,ms\prime>$ that represent the state with
one electron of spin $s$ at site $n$ and the other with spin
$s\prime$ at site $m$:

\begin{equation}
\Phi _{E}=\sum_{ns,ms^{\prime }}g(ns,ms^{\prime };E)\mid ns,ms^{\prime }>
\label{phie}
\end{equation}

In the Wannier representation we obtain the following set of
equations corresponding to energy $E$:

\begin{widetext}
\begin{equation}
V\,\,(g_{n+1,m}+g_{n,m+1}+g_{n-1,m}+g_{n,m-1})\,+\,(\gamma _{n}+\gamma _{m}+U\delta
_{n,m})\,\,g_{n,m}\,=\,E\,g_{n,m}  \label{vg}
\end{equation}
\end{widetext}

In this equation the first index refers to a particle with spin $up$
and the
second for spin $down$. The Wannier amplitudes $g_{n\uparrow ,m\downarrow }$%
do not depend on time. For simplicity we omit the label $E$ in the Wannier
amplitudes.

For the crystalline case and without the presence of an electric field, all $%
\gamma _{i}$ can be taken as zero. Introducing the center of mass and
relative coordinates in units of the lattice parameter:

\begin{equation}
R=(n+m)/2; \hspace{1.5cm} r=n-m  \label{r}
\end{equation}
and following Hubbard we expand the Wannier amplitudes:

\begin{equation}
g_{n,m}=\sum_{K,k}\Phi (K,k)\,\,e^{iKR}e^{ikr}  \label{g}
\end{equation}
to obtain the following equation for the $\Phi (K,k):$

\begin{equation}
\Phi (K,k)=\frac{\frac{U}{N}\text{ }\sum_{k^{\prime }}\Phi (K,k^{\prime })}{%
E-2V\cos (\frac{K}{2}+k)-2V\cos (\frac{K}{2}-k)}  \label{phik}
\end{equation}
which in turn becomes:

\begin{equation}
1=\frac{U}{N}\sum_{k}\frac{1}{E-\varepsilon_{\frac{K}{2}+k}-
\varepsilon_{\frac{K}{2}-k}}  \label{e1}
\end{equation}

\begin{figure}[htbp]
\centerline{\includegraphics[width=6.5cm,angle=-90]{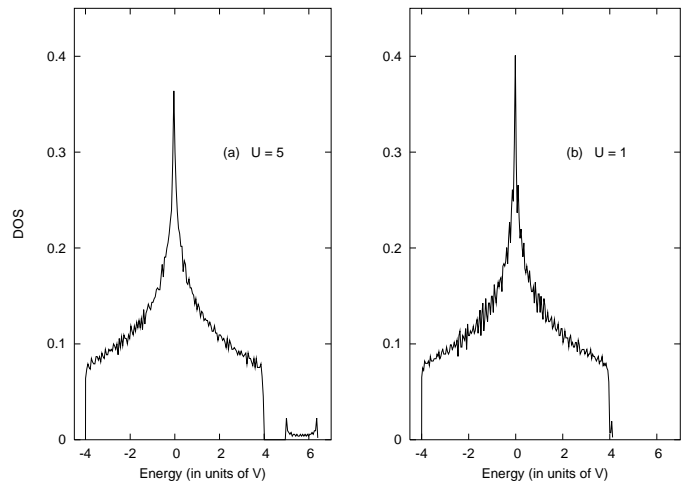}}
\caption{The density of states in a crystal for: a) $U = 5$, and b)
$U = 1$. The energy is in units of the hopping parameter V.}
\label{dos}
\end{figure}

For every $K$ we obtain $N-1$ roots of this equation inside the
infinities of the right hand side for which $E= \,\varepsilon
_{\frac{K}{2}+k}+ \varepsilon_{\frac{K}{2}-k}\,=\,4V\cos
\frac{K}{2}\cos k$. At the same time, and provided $U$ is large
enough, there appears an extra root outside the band. In this
case, by varying $K$ we get an excited second band. By looking at
the eigenvalue equation~(\ref{vg}) we note its close resemblance
with the corresponding Schr\"{o}dinger equation for \textit{one}
particle in a tight-binding $2D$ Hamiltonian, except for the $U$
Hubbard term~\cite{lp}. The obtained density of states for the
crystalline case for $U=5V$ shown in Fig.~\ref{dos}.a, present two
bands. The lower band is very similar to the case of one electron
in $2D$ for the reasons stated above and, the excited band has the
typical structure of tight-binding $1D$ bands. Moreover, by
looking at the equation~(\ref{e1}) one notes that $E=U$ is a root
of it, which happens for $K=\pm \pi .$ This way if $U$ is less
than $4V$ we obtain resonant states inside the band, instead of an
excited band separated from the $2D$ band for a $U-4V$ gap, that
appears when $U$ is greater than $4V$, since the bottom of the
excited band occurs at $E=U$. See Fig.~\ref{dos}.b. We conclude
that the inclusion of the electron-electron interaction in a
crystalline structure in $1D$ have produced a band with a
structure similar to a $2D$ band for \textit{one} electron, plus
an extra band provided $U$ is great enough.

\section{The dynamics}

We follow the time evolution of wave packets in a $1D$ lattice, by assuming
the following expansion for the wave function in the Wannier representation

\begin{equation}
\mid \Psi (t)>\,\,=\,\sum_{ns,ms^{\prime }}f_{ns,ms^{\prime}}(t)\,\mid
ns,ms^{\prime }>  \label{psi}
\end{equation}
where the Wannier amplitudes now depend on time. The time evolution of the
wave function

\begin{equation}
i\hbar \,\,\frac{\partial \mid \Psi >}{\partial t}\,\,=\,H\mid \Psi >  \label{ham1}
\end{equation}
becomes in the Wannier representation:

\begin{widetext}
\begin{equation}
i\hbar \,\,\frac{df_{n,m}}{dt}\,=\,V\,\,(f_{n+1,m}+f_{n,m+1}+f_{n-1,m}+f_{n,m-1})
\,+\,(\gamma _{n}+\gamma _{m}+U\delta _{n,m})\,\,f_{n,m}  \label{df}
\end{equation}
\end{widetext}

We assume the initial condition with an electron around site $p$ with spin $%
up$ and the other around site $q$ with spin $down$. In order to do this, we
shall consider the initial condition described by gaussian packets of
different standard deviations centered at sites $p$ and $q$:

\begin{equation}
\mid \Psi (t=0)>\,=\,\mid \mathbf{p}_{\mathbf{\uparrow }}\mathbf{,q}_{%
\mathbf{\downarrow }}>  \label{psi1}
\end{equation}
where

\begin{equation}
<n\mid \mathbf{p}_{\mathbf{\uparrow }}(0)>\,=\,C\exp \biggl[-\frac{(n-p)^{2}%
}{2\sigma ^{2}}\biggr]  \label{n}
\end{equation}

\begin{equation}
<m\mid \mathbf{q}_{\mathbf{\downarrow }}(0)>\,=\,C\exp \biggl[-\frac{%
(m-q)^{2}}{2\sigma ^{2}}\biggr]  \label{m}
\end{equation}

\noindent and $C$ is a normalization constant.

As said before, all $\gamma _{i}$ can be taken as zero for the crystalline
case without an electric field. We shall consider gaussians with standard
deviations $\sigma$. The case $\sigma =0$ corresponds to a particle
localized in a single site. In the crystalline case where all $\varepsilon
_{n}$ are the same and not taking into account the Hubbard electron-electron
interaction, both particles should propagate ballistically in the regular
lattice~\cite{lp}. But now, in the presence of the on-site interaction $U$,
a different behavior should take place. To follow the evolution of the
particles injected around sites $p$ and $q$ with spins up and down
respectively, we perform the following. We define the mean-square
displacement of each electron as indicated below.

\begin{equation}
MSD_{p\uparrow}\,=\,\sum_{n}(n-p)^{2}\,\,\sum_{m}\mid f_{n,m}(t)\mid ^{2}
\label{msd}
\end{equation}

\noindent and

\begin{equation}
MSD_{q\downarrow}\,=\,\sum_{m}(m-q)^{2}\,\,\sum_{n}\mid f_{n,m}(t)\mid ^{2}
\label{msd1}
\end{equation}

At the same time we can follow the centroid of the wave packet associated
with the movement of each electron as follows:

\begin{equation}
<\Delta x(t)>_{p\uparrow }\,\,=\,\sum_{n}\,(n-p)\,\sum_{m}\,\mid f_{n,m}(t)\mid ^{2}
\label{xt}
\end{equation}
and

\begin{equation}
<\Delta x(t)>_{q\downarrow }\,\,=\,\sum_{m}\,(m-q)\,\sum_{n}\,\mid f_{n,m}(t)\mid ^{2}
\label{xt1}
\end{equation}
which will give us the amount of the displacement from the initial positions
for both electrons.

At the same time, by taking the absolute value of the difference
between the respective centroid positions, we can determine the
time evolution of the mean distance of the two electrons, or the
size of the propagating pair: $d(t)\,=\,\,\, \mid
\,<x(t)>_{p\uparrow }-\text{ }<x(t)>_{q\downarrow}\,\mid$.

%\begin{equation}
%d(t)\,=\,\,\, \mid \,<x(t)>_{p\uparrow }-\text{
%}<x(t)>_{q\downarrow}\,\mid
%\end{equation}

In what follows we shall define the energies in units of the
hopping parameter $V$, and the time in dimensionless units $\tau
\,=\,Vt / \hbar$.

\section{Interplay between U and the initial wave
packet extension for a Crystalline case}

\begin{figure}[htbp]
\centerline{\includegraphics[width=6.5cm,angle=-90]{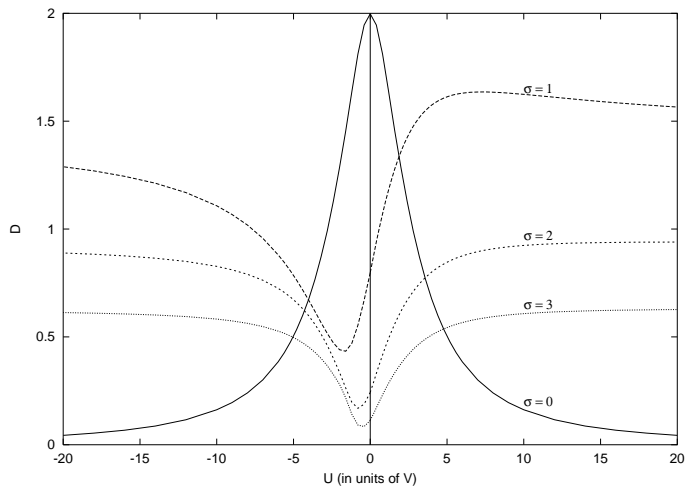}}
\caption{The MSD coefficient $D$ as function of the interaction $U$
for four cases of initial wave packets: $\sigma = 0$, $1$, $2$ and
$3$. As said in the text $MSD/,=/,D/,t^2+C$} \label{coeffs}
\end{figure}

In order to study the combined effect that the interaction and the
extension of the initial wave packet have on the propagating
properties, we shall analyze the behavior of the MSD's as function
of time. We consider the case in which both particles are injected
at $t=0$ in the same position in the lattice. By taking four
different initial waves associated with $\sigma = 0$, $1$, $2$ and
$3$ we note ballistic propagation in all cases, i.e., $MSD =
Dt^2+C$. In Fig.~\ref{coeffs} we have plotted the ballistic
coefficient D as function of the interacting Hubbard term U.

We treat first the case of non-interacting electrons, i.e., $U=0$.
We note that the MSD's values decrease when the extension of the
initial packets increase. The broader the packet the more inertia
it has, it becomes more "massive".

We consider now the interacting case. First,  we concentrate on
the $\sigma = 0$ case and vary the strength of the Hubbard
potential. The U term acts as an on-site energy so that by
increasing it, we introduce a barrier that inhibits hopping, the
more so, the bigger U is. We still have propagation but with
smaller MSD's values as we increase U. The ballistic coefficient
$D$ is a monotonically decreasing function of U.

An interesting situation occurs when the initial wave packet has
an extension, i.e., $\sigma \neq 0$. At this point we should state
that the hopping parameter was assumed to be negative. By looking
at the dynamical equation (10) we realize that what is relevant in
analyzing propagation is the difference U - V, which can be
considered like an effective interaction. This way, when both
magnitudes have the same sign the interaction is reduced, while
when they have different sign the effective interaction is
increased. Consider first the case $U > 0$. We obtain in all cases
studied ($\sigma$ = $1$, $2$ and $3$) the D coefficient increases
rapidly with U until reaching a plateau while $D(\sigma_1)>
D(\sigma_2) > D(\sigma_3)$, i.e., the more extended the initial
wave packet, the more massive it is. This also happens for the non
interacting case $U = 0$, as shown above.

Now for $U < 0$ a different behavior is observed, where in this
case the effective interaction is reduced. First we take $\sigma =
1$ and see that for $U = -1$ {\it smaller} values than for $U = 0$
are obtained, while for U= -2 we get even smaller values than for
$U = -1$. From then on the MSD's increase monotonically with $\mid
U \mid$, such that from $\mid U \mid = 5$ on, the values are
greater than the corresponding to $U = 0$. The minimum of the
coefficient D occurs for $U = -1.5$. Taking a more extended
initial wave packet, $\sigma = 2$, the same trend is observed, but
now the minimum is at $ U = -0.75 $. Already for $U = -2$ the MSD
is larger than for $U = 0$, and from then on they increase
monotonically with $\mid U \mid$, as it happens for $\sigma = 1$.
As for $\sigma = 3$ it shows the same behavior with the minimum
closer to U = 0.

We will explain now the initial decreasing trend for small $\mid U
\mid$, when the initial packet has extension different from zero
and hopping and interaction parameters have same sign.  We have
two competing effects to consider, on one side the extension of
the packet implies more inertia (smaller MSD values) as said
above, on the other hand, the presence of the U repulsive term
acts on the different components of the wave trying to push the
packet further away. For small $\mid U \mid$, the first effect
overcome so one gets smaller MSD values but, increasing $\mid U
\mid$, the repulsion dominates so greater values are obtained.
Finally, we note a cross over region that happens for  $4 <
\,\,\mid U \mid \,\,< 5$, where for the three cases considered:
$\sigma = 1$, $2$ and $3$, the corresponding MSD values are
greater than the ones for $\sigma = 0$, for $\mid U \mid \,\gtrsim
5$.

\begin{figure}[htbp]
\centerline{\includegraphics[width=8cm,angle=0]{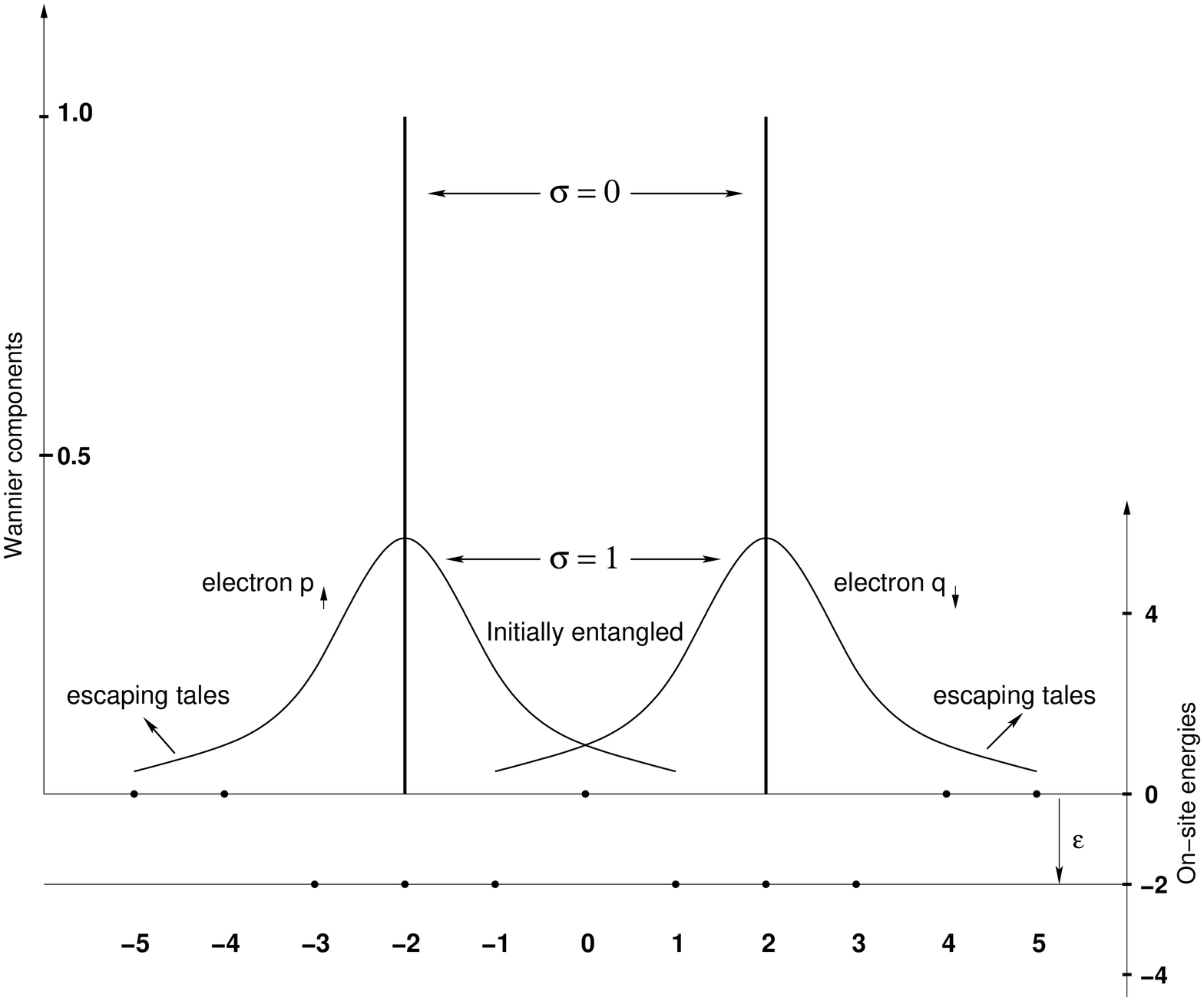}}
\caption{The initial wave packets for the two configurations
assumed: well localized ($\sigma = 0$) and extended ($\sigma =
1$).}\label{initial}
\end{figure}

\section{Two electrons in a crystal with impurities}

 We shall consider injected particles at time zero around
sites $p$ and $q$, where these sites as well as their nearest
neighbors are taken as impurity levels in an otherwise regular
lattice. In Fig.~\ref{initial}, we show the initial wave packets
for the configurations $\sigma = 0$ and $1$. We illustrate the
cases in which the packets are centered at positions $\pm 2$, and
the impurity levels at $\varepsilon =-2$, see Fig.~\ref{initial}.
Note that for the case $\sigma = 1$ the initial wave packets
overlap due to the vicinity of the sites $p$ and $q$. Obviously
for $\sigma = 0$ no overlap is obtained.

\begin{figure}[htbp]
\centerline{\includegraphics[width=6cm,angle=-90]{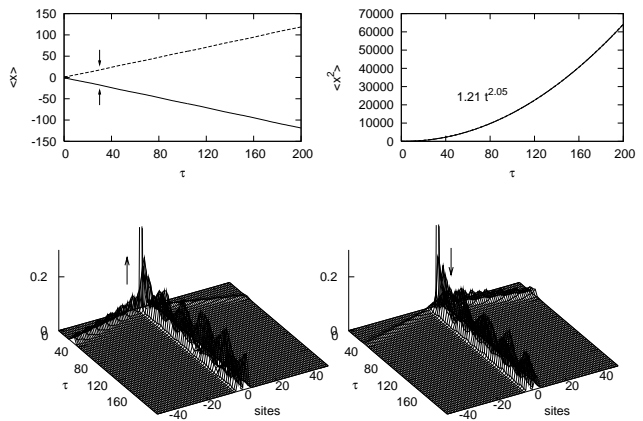}}
\caption{Show for $\varepsilon =-2V$, $U = 4$ and $\sigma = 0$, in
the upper part, the centroids and the MSD as functions of time and
in the lower part the time evolution of the up and down spin wave
packets. Note that a considerable amount of both waves get trapped
around the starting positions, while the rest propagate rapidly
given rise to the observed superballistic behavior. }
\label{e-2U4s0}
\end{figure}

\begin{figure}[htbp]
\centerline{\includegraphics[width=6cm,angle=-90]{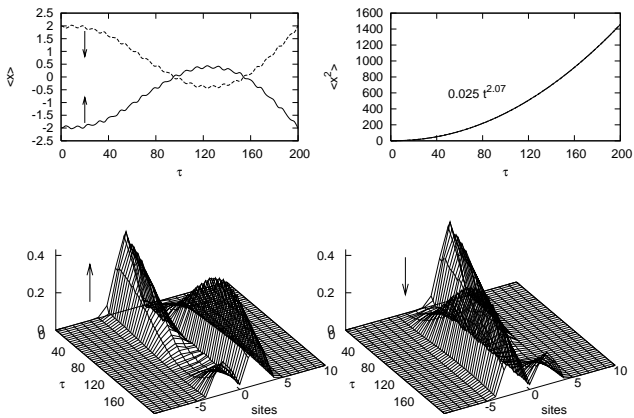}}
\caption{The same as in Fig.~\ref{e-2U4s0} but for $\varepsilon
=-2V$, $U = 4$ and $\sigma = 1$. In this case the more extended
initial wave packets overlap given rise as a result to an
entanglement of the two particles that perform oscillatory movements
as time goes. Note that the MSD values are much smaller than in the
previous case $\sigma = 0$.} \label{e-2U4s1}
\end{figure}

First we assumed the impurity energy levels at $\varepsilon =-2$
and varied $\sigma$ as well as the interacting factor U. We shall
discuss the results for the initially well localized wave packets
($\sigma = 0$) and moderate interaction, $U = 4$, see
fig.~\ref{e-2U4s0}. The centroids clearly start to depart from
each other as soon as the up and down packets overlap due to the
interaction. The MSD shows a superballistic behavior: $Ct^\alpha$
with a time exponent $\alpha = 2.05$. One notes by looking at the
wave packets evolution that rather big components of the two waves
get trapped around the starting positions due to the well depth.
At the same time, the tails of the initial packets are able to
propagate, being responsible for the superballistic behavior.

Now by taken $\sigma = 1$ we obtain a completely different
picture, we observe an oscillatory movement of the centroids, see
Fig.~\ref{e-2U4s1}. The wave corresponding to each of the
particles get trapped in a very small region and one notices that
they strongly overlap in such a way that when one is moving to the
right the other perform a displacement to the left and vice versa.
The up and down spin wave packets get entangled with each other.
The trapping of the wave packets can be understood by noticing
that the sites covered by the gaussians are degenerate with
on-site energies very different from the rest of the lattice, thus
inhibiting hopping to distant sites. The behavior of the MSD's
proportional to $t^2$ is due to the escaping of the tails of the
initial wave packets that can hop to the near degenerate sites.
This case serve to indicate that the propagating properties can
not be inferred only from the MSD. In this connection we resort to
3D graphs that show the way the packets evolve in time.

\begin{figure}[htbp]
\centerline{\includegraphics[width=6cm,angle=-90]{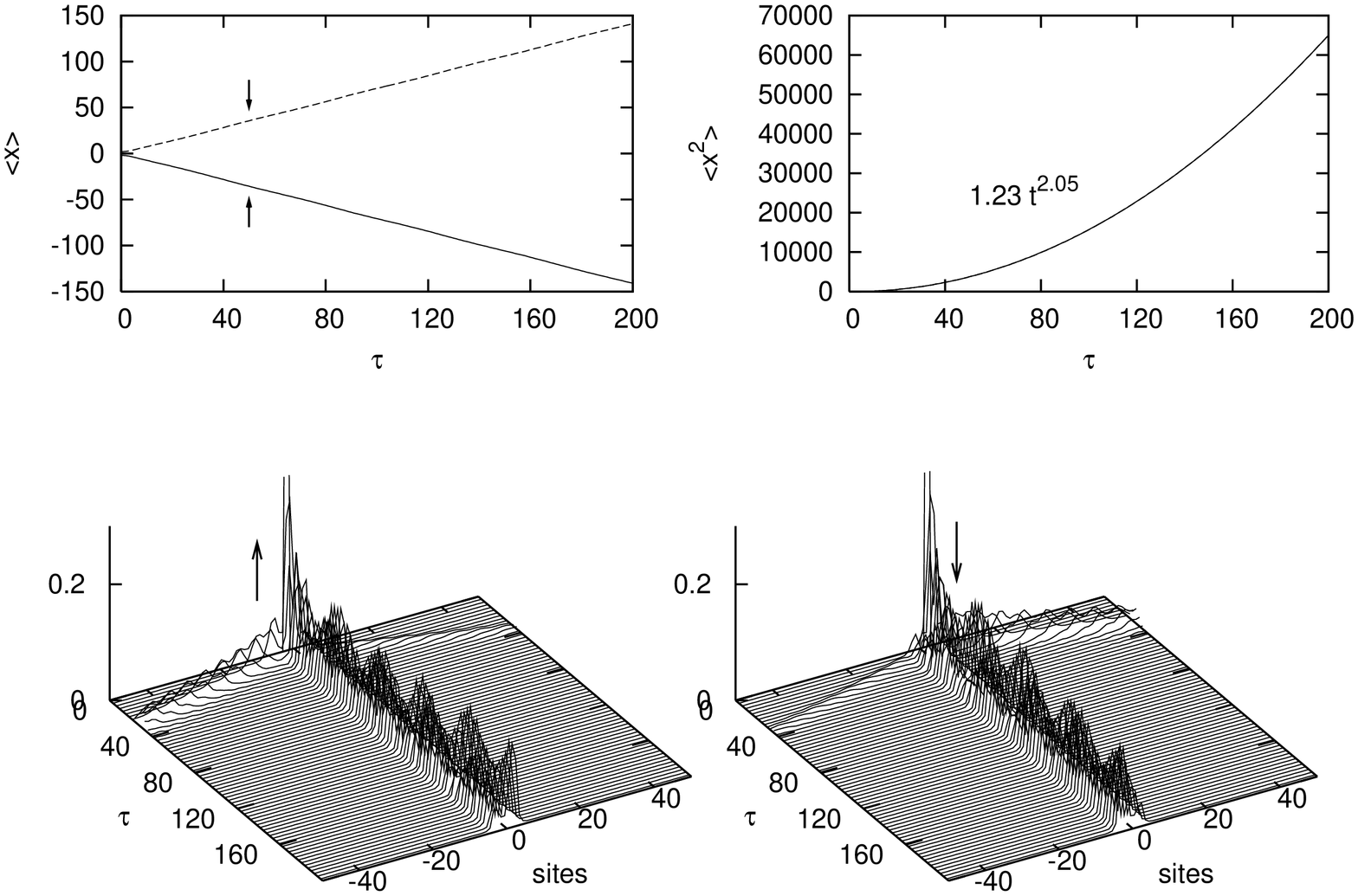}}
\caption{The same as in Fig.~\ref{e-2U4s0} but for $\varepsilon
=-2V$, $U = 8$ and $\sigma = 0$}. \label{e-2U8s0}
\end{figure}

\begin{figure}[htbp]
\centerline{\includegraphics[width=6cm,angle=-90]{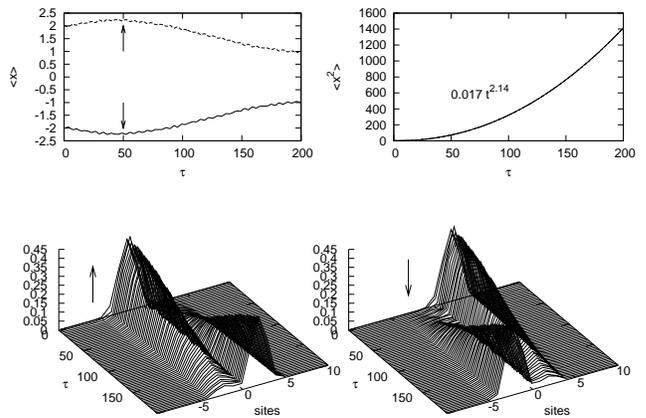}}
\caption{The same as in Fig.~\ref{e-2U4s0} but for $\varepsilon
=-2V$, $U = 8$ and $\sigma = 1$. Note the analogous behavior as the
shown in the case for $U = 4$.} \label{e-2U8s1}
\end{figure}

Increasing the interaction factor to $U = 8$ and taken $\sigma =
0$ we obtain a similar behavior than the one described above for
$U = 4$, see Fig.~\ref{e-2U8s0}. For the same U and $\sigma = 1$
we observe the same entanglement that occurs for $U = 4$, with the
difference that the period of the oscillation of the centroids is
larger in the present case, see Fig.~\ref{e-2U8s1}.

\begin{figure}[htbp]
\centerline{\includegraphics[width=6cm,angle=-90]{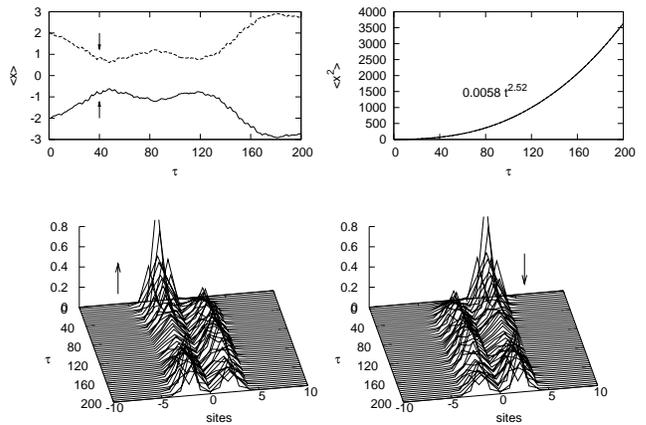}}
\caption{The same as in Fig.~\ref{e-2U4s0} but for $\varepsilon
=-4V$, $U = 4$ and $\sigma = 0$.} \label{e-4U4s0}
\end{figure}

\begin{figure}[htbp]
\centerline{\includegraphics[width=6cm,angle=-90]{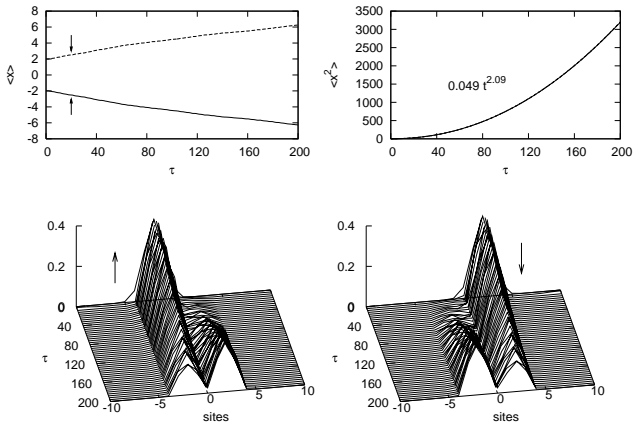}}
\caption{The same as in Fig.~\ref{e-2U4s0} but for $\varepsilon
=-4V$, $U = 4$ and $\sigma = 1$.} \label{e-4U4s1}
\end{figure}

We treat now the case in which the impurity levels are deeper,
i.e., at $\varepsilon =-4$. For the initially well localized
packets, $\sigma = 0$, and the interaction $U = 4$, we note that
the centroids tend to get near to each other until they fill the
repulsion, but they do not depart very much and later on they
repeat the tendency to get closer. As for the MSD's the values are
much smaller than the corresponding for $\varepsilon =-2$. By
looking at the evolution of the wave packets in the 3D graphs it
is evident that the packets get trapped due to the deepness of the
wells, see Fig.~\ref{e-4U4s0}. Considering now  more extended
initial wave packets, $\sigma = 1$, we note that the centroids
depart from each other, the MSD's show ballistic behavior due to
the escaping tales of the initial wave packets. As for the time
evolution of the up and down wave functions we observe that they
get trapped around the starting positions. A small amount of the
amplitude of each of the up and down waves show a tendency to
overlap with the other, see Fig.~\ref{e-4U4s1}.

\begin{figure}[htbp]
\centerline{\includegraphics[width=6cm,angle=-90]{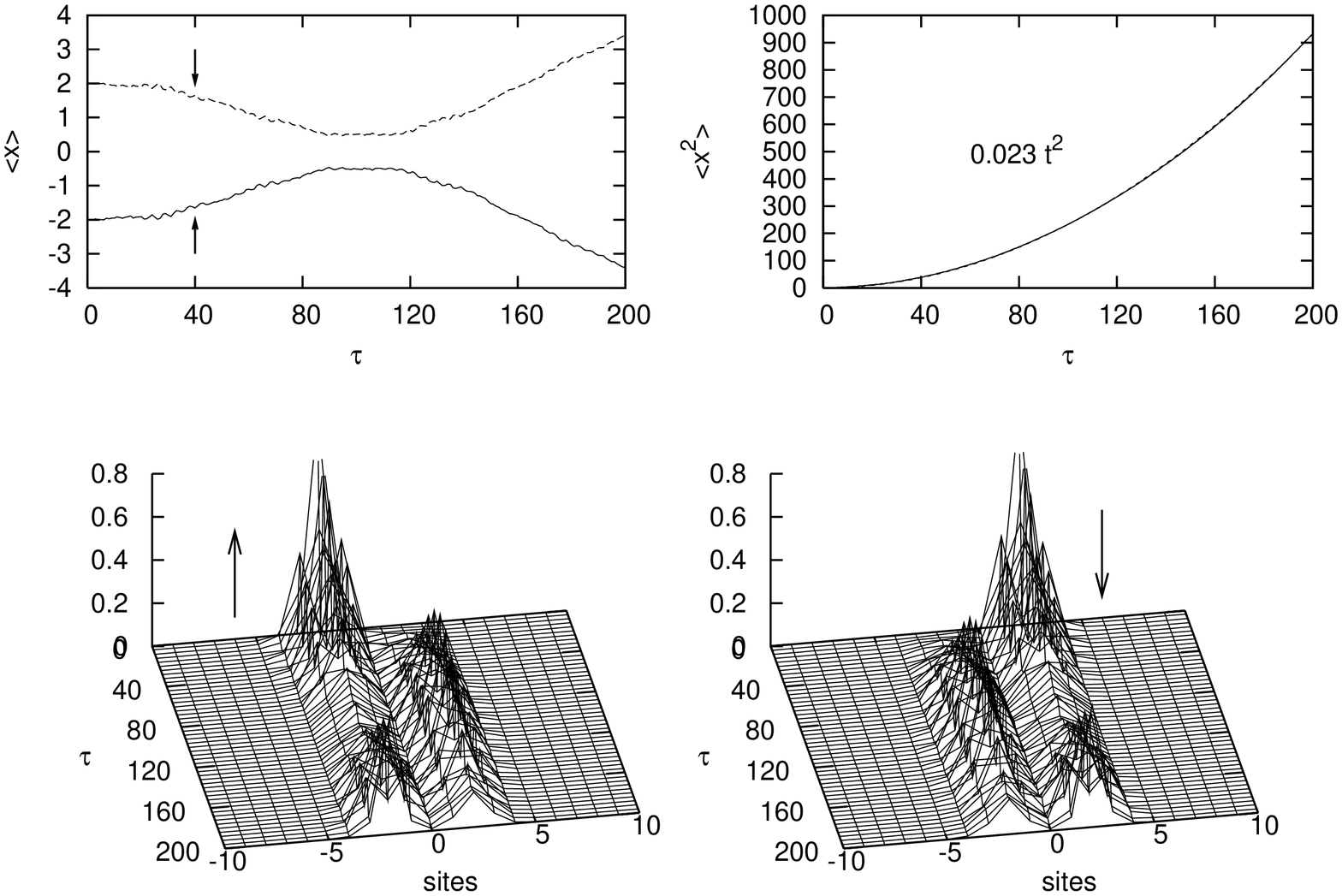}}
\caption{The same as in Fig.~\ref{e-2U4s0} but for$\varepsilon
=-4V$, $U = 8$ and $\sigma = 0$.} \label{e-4U8s0}
\end{figure}

\begin{figure}[htbp]
\centerline{\includegraphics[width=6cm,angle=-90]{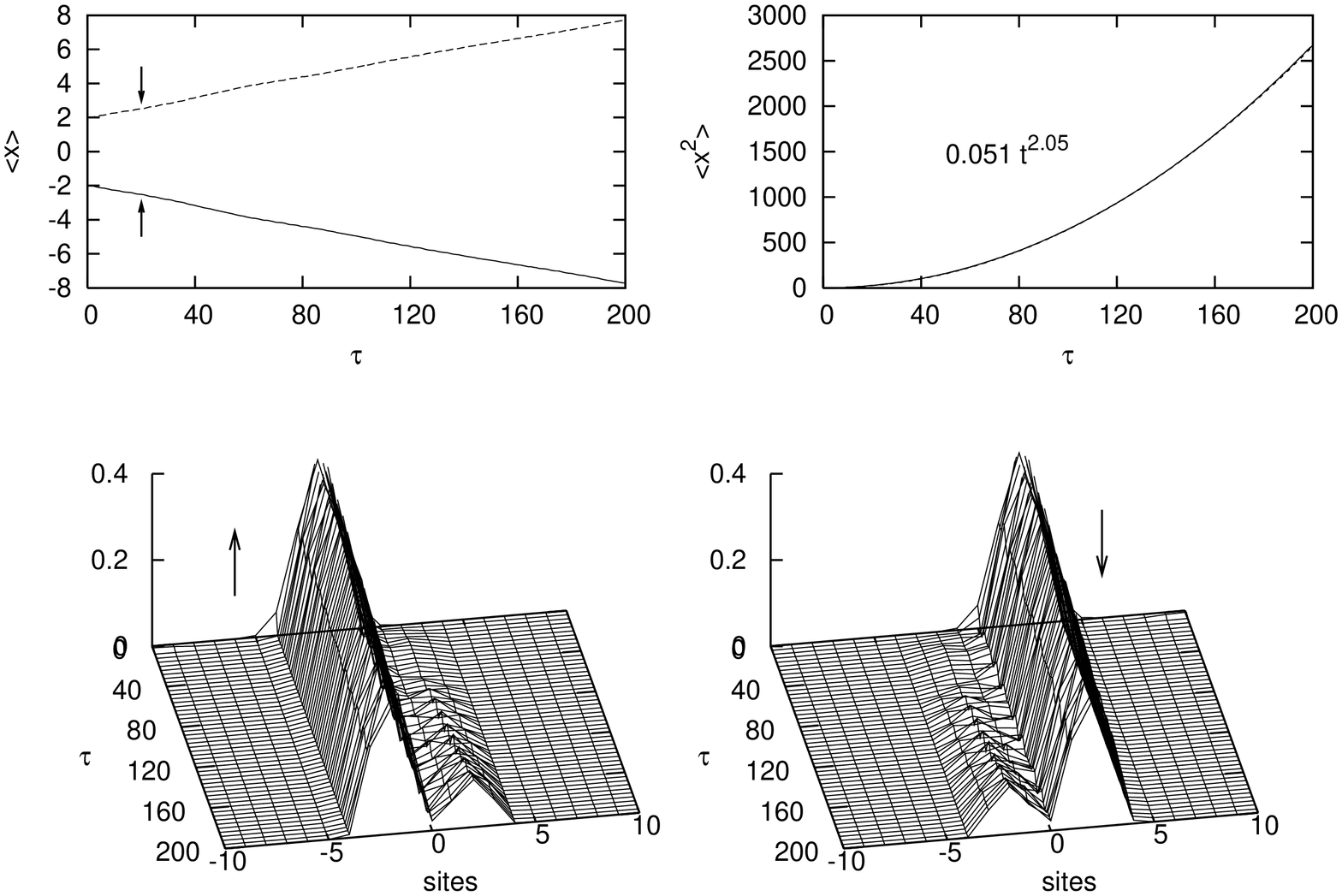}}
\caption{The same as in Fig.~\ref{e-2U4s0} but for$\varepsilon
=-4V$, $U = 8$ and $\sigma = 1$.}\label{e-4U8s1}
\end{figure}

Now for a strong interaction, $U = 8$ and taken $\sigma = 0$, we
observe that the centroids tend to approximate while the MSD's
values are much smaller than the corresponding to $\varepsilon
=-2$. Due to the deeper impurities levels, the packets get trapped
in a very small region of the lattice, see Fig.~\ref{e-4U8s0}.
When we take more extended initial packets, i.e., $\sigma = 1$,
the centroids tend to depart from each other, the MSD's are larger
than the corresponding to the former case, $\sigma = 0$, and the
wave packets perform the same kind of oscillations, see
Fig.~\ref{e-4U8s1}.

The time limit taken in our calculation was $10^{-11}$sec, longer
than any reasonable collision time, this implies that in order to
avoid undesirable boundary effects, we have included lattices up
to 1000 sites. We used the Runge-Kutta method to integrate the
equations of motion.

\section{Conclusions}

In this work we analyzed the role the electron-electron
interaction plays in one-dimensional disordered systems, in
particular, a crystalline lattice with impurities. Two electrons
with opposite spins were introduced in a band where their
interaction was assumed along the Hubbard model. We have obtained
the energy spectrum of a crystal considering different values of
the interaction strength. The obtained density of states show that
for sufficiently large $U$, two bands are present separated by a
gap of intensity $U - 4V$. The lower band has the feature of the
one corresponding to a single particle in a 2D crystal, while the
excited band has the structure of a 1D tight-binding crystal.

As for the dynamical properties, we analyzed the interplay between
the initial wave packet extensions and, for both, positive and
negative values of the interaction $U$.

In the non-interacting case, $U = 0$, we show that by increasing
the extension of the packets, one reduces the MSD values, the more
so, the more extended the initial wave packets, they become more
"massive".

In the case of a positive U, the obtained MSD increases rapidly
from the non interacting case $U = 0$. For negative $U$, a
different initial behavior is observed. As $U$ is near the origin
the MSD decrease until a minimum value is achieved, and from then
on it increases reaching a plateau. For both the positive and
negative values of $U$ there exists a cross over region as
explained in the text.

Next we analyze the propagation that takes place when we introduce
the two interacting particles at impurity sites in an otherwise
periodic crystal. We have encountered situations for which the MSD
as function of time presents superballistic behavior, although
when looking at the packet we realize that most of it get trapped
in a small region of the lattice. That is why the propagating
properties can not be inferred from the MSD alone. When we
consider moderate impurity levels, in our case $\varepsilon =-2V$,
the response of the system with regard to the initial conditions
are very different. In fact, when we start with very well
localized particles, $\sigma = 0$, the centroids depart from each
other for every intensity of the interaction, while for the less
localized initial packets, $\sigma = 1$, the two particles get
entangled, trapped in the impurity region, we observe the
centroids performing a periodic movement. When considering deeper
impurity levels such as $\varepsilon =-4$ no matter what the
extension of the initial waves packets are, the result indicate
that the particles remain localized around the starting positions,
this is true for every interaction strength.

%We studied two different situations with regard to the starting
%positions of the particles: i) both start at the same site and ii)
%they start at different positions in the lattice. In the former
%case and due to the nature of the on-site Hubbard term, the two
%particles remain together as a single entity. For the latter case,
%the $U$ term plays an important role, since its inclusion in the
%equations produces a departure of the centroids associated with
%each particle, that is, it has a repulsive effect on the
%trajectories. This effect is present for every system as long as
%the initial wave packets do not overlap. In the other hand, when
%the two particles are injected at impurity sites in an otherwise
%periodic structure and the initial packets associated with each of
%them overlap, we observe an oscillatory movement of the centroids
%trapped in the impurity region, the two particles get entangled
%with each other.

\end{document}